\documentclass[twocolumn,showpacs,amsmath,amssymb,nofootinbib]{revtex4}
\usepackage{epsfig}
\usepackage{bm}
\usepackage[usenames]{color}

\begin{document}

\title{On resonance search in dilepton events at the LHC}

\author{M. V. Chizhov$^{1,2}$, V. A. Bednyakov$^1$, J. A. Budagov$^1$}
\affiliation{$^{\it 1}$Dzhelepov Laboratory of Nuclear Problems,\\
\mbox{Joint Institute for Nuclear Research, 141980, Dubna,
Russia}\\
$^{\it 2}$Centre for Space Research and Technologies, Faculty of
Physics, Sofia University, 1164 Sofia, Bulgaria}


\begin{abstract}
The main distribution for a bump search is the dilepton invariant
mass distribution with appropriated cut on an absolute value of
pseudorapidity difference $\Delta \eta\equiv|\eta_1-\eta_2|$ between
the two leptons. The background from the Standard Model Drell--Yan
process contributes mainly to the central pseudorapidity region
$\Delta\eta\simeq 0$. By contrast, the excited bosons lead to a peak
at $\Delta\eta\simeq 1.76$. We show that this property allows to
enhance the significance of their bump search by means of new cut
optimization. Nevertheless, in order to confirm an observation of
the bump and reveal the resonance nature other angular distributions
should be used in addition.
\end{abstract}


\maketitle

\section{Introduction}

The golden channel of resonances search at the hadron colliders is
the lepton one with electrons or muons in a final state. Free from
huge QCD background it opens clean way of the resonance
identification. Depending on the resonance electric charge there are
two possible signatures of the resonances search through their decay
into a charged lepton and a neutrino with missing transverse
momentum or into a lepton pair (Drell--Yan production process). The
latter has the very clean signature and almost totally
reconstructible kinematics\footnote{Up to small individual
transverse momenta of the quarks.}, which allows investigation of
the production and the decay of the new resonances.

In this letter we will consider the case of the resonance production
of new bosons up to spin 2 and their detection in the Drell--Yan
process using the first CERN LHC data. We will concentrate mainly on
the case of excited bosons, which have unique angular distribution.
The simple model of the excited bosons was proposed in
\cite{proposal}. Roughly speaking it leads to a signal, $s$, of the
two orders of magnitude higher than the Standard Model (SM)
background, $b$. Therefore, as for the other benchmark models, the
direct exclusion limits on mass of 1.152~TeV and on cross section
times branching ratio $\sigma B(Z^*\to\ell^+\ell^-) < 89$~fb have
been obtained for the first time, based on 2010 ATLAS
data~\cite{ATLAS}.

In general case the coupling constants of the new bosons with quarks
and leptons could be weaker and their masses could be lighter than
the obtained limit. In this case the signal significance,
$s/\sqrt{b}$, will decrease, which lead to more complicated
situation in the resonance search. The main background comes from
the inclusive Drell--Yan process $pp\to \gamma/Z\to\ell^+\ell^-$
with $\gamma/Z$ exchanges and is practically unremovable. Therefore,
instead to wait for more data one can involve angular distributions
into analysis~\cite{angular} as well.

We will use unique properties of angular distributions of the
excited bosons in order to increase their signal sensitivity in the
present data. In the next section we will consider a model
independent signal distributions of new physics from resonances up
to spin 2. Present analysis is based on our study of dijet mass
angular distributions~\cite{dijets}, which has been shown to play
important role in disentangling of the resonance properties and
revealing the unique signature of the excited bosons.

\section{A unique signal of excited bosons}

The distribution of final state lepton over the polar scattering
angle $\theta$, being angle between the axis of the lepton pair and
the beam direction in the dilepton rest frame, is directly sensitive
to the dynamics of the underlying process, where the spins of the
resonance and of the initial and final states uniquely define the
angular distribution.

In the following we will consider a distribution on the absolute
value of the dilepton pseudorapidity difference, which is related to
the angle $\theta$ by $\Delta\eta\equiv
|\eta_1-\eta_2|=\ln[(1+|\cos\theta|)/(1-|\cos\theta|)]\ge 0$. It has
been shown~\cite{deta} that the excited boson distribution on
$\Delta\eta$ drastically differ from the SM and other exotic models.
So, while the SM processes are dominated by $\gamma/Z$ exchanges,
which lead to well-known $1+\cos^2\theta$ distribution\footnote{The
choice of the variables, which depend on the absolute value of
$\cos\theta$, cancels out the apparent dependence on $\cos\theta$.},
exotic physics processes can deviate from this distribution.

Let us consider different possibilities for the spin of a resonance
and its possible interactions with quarks and leptons. The simplest
case of the resonance production of a (pseudo)scalar particle $h$
with spin 0 in $s$-channel leads to a uniform decay distribution on
the scattering angle
\begin{equation}\label{G0}
    \frac{{\rm d} \Gamma_0(h\to \ell^+\ell^-)}
    {{\rm d} \cos\theta} \propto \vert d^{\,0}_{00}\vert\,^2\sim 1.
\end{equation}

\begin{figure}[th]
\epsfig{file=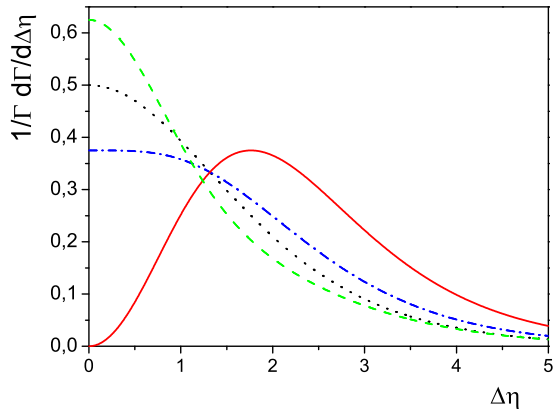,width=0.5\textwidth}
\caption{\label{fig:angular} The normalized angular lepton
distributions as functions of $\Delta\eta$ for the scalar, spin-1
bosons with the minimal couplings, the excited bosons and spin-2
resonances are shown with the dotted, dash-dotted, solid and dashed
curves, respectively.}
\end{figure}

Variable transformation
\begin{equation}\label{trans}
    \frac{{\rm d}\Gamma}{{\rm d}\Delta\eta}=
    \frac{{\rm d}\cos\theta}{{\rm d}\Delta\eta}\;
    \frac{{\rm d}\Gamma}{{\rm d}\cos\theta}\,,
\end{equation}
in the uniform distribution (\ref{G0}) from $\cos\theta$ to
$\Delta\eta$
\begin{equation}\label{dy0}
    \frac{1}{\Gamma_0}\frac{{\rm d}\Gamma_0}{{\rm d}\Delta\eta}=
    \frac{2{\rm e}^{\Delta\eta}}{({\rm e}^{\Delta\eta}+1)^2}
\end{equation}
leads to kinematical peak at $\Delta\eta=0$ (the dotted curve in
Fig.~\ref{fig:angular}), which corresponds to the polar angle
$\theta=90^\circ$. The result coincides with common opinion, that an
expected signal of new physics should be in perpendicular direction
to the beam.

According to \cite{dijets,Gia} there are only two different
possibilities for the spin-1 resonances to interact with light
fermions. All known gauge bosons have minimal interactions with the
light fermions
\begin{equation}\label{LL}
    {\cal L}_{Z'}=\sum_f \left(g^f_{LL}\;\overline{\psi^f_L}\gamma^\mu\psi^f_L
    +g^f_{RR}\;\overline{\psi^f_R}\gamma^\mu\psi^f_R\right) Z'_\mu \,
    ,
\end{equation}
which preserve the fermion chiralities and possess maximal
helicities $\lambda=\pm 1$. It is assumed that the hypothetical $Z'$
bosons have also similar couplings. At a symmetric $pp$ collider,
like the LHC, such interactions lead to the specific symmetric
angular distribution of the resonance decay products
\begin{equation}\label{GLL}
    \frac{{\rm d} \Gamma_1(Z'\to\ell^+\ell^-)}
    {{\rm d} \cos\theta} \propto
    \vert d^1_{11}\vert^2+\vert d^1_{-11}\vert^2 \sim
    1+\cos^2\theta \, .
\end{equation}
Similar to the uniform distribution, eq.~(\ref{GLL}) also leads to
kinematical peaks at $\Delta\eta=0$  (the dash-dotted curve in
Fig.~\ref{fig:angular}):
\begin{equation}\label{dy1prime}
    \frac{1}{\Gamma'_1}\frac{{\rm d}\Gamma'_1}{{\rm d}\Delta\eta}=
    \frac{3{\rm e}^{\Delta\eta}({\rm e}^{2\Delta\eta}+1)}{({\rm e}^{\Delta\eta}+1)^4}.
\end{equation}

Another possibility is the resonance production and decay of new
longitudinal spin-1 $Z^*$ bosons with helicity $\lambda=0$. The new
gauge bosons with such properties arise in many
extensions~\cite{Gia} of the SM, which solve the Hierarchy Problem.

While the $Z'$ bosons with helicities $\lambda=\pm 1$ are produced
in left(right)-handed quark and right(left)-handed antiquark fusion,
the longitudinal $Z^*$ bosons are produced through the anomalous
chiral couplings with the ordinary light fermions
\begin{equation}\label{LR}
    {\cal L}_{Z^*}=\sum_{f}
    \frac{g^f_{RL}}{M}\,\overline{\psi^f_R}\sigma^{\mu\nu}
    \psi^f_L\;\partial_{[\mu} Z^*_{\nu]}+{\rm h.c.}
\end{equation}
in left-handed or right-handed quark-antiquark
fusion~\cite{proposal}. The anomalous interactions (\ref{LR}) are
generated on the level of the quantum loop corrections and can be
considered as effective interactions. The $Z^*$ resonances are some
types of ``excited'' states as far as the only orbital angular
momentum with $L=1$ contributes to the total angular moment, while
the total spin of the system is zero. This property manifests itself
in their derivative couplings to fermions and a different chiral
structure of the interactions in contrast to the minimal gauge
interactions (\ref{LL}).

The anomalous couplings lead to a different angular distribution
\begin{equation}\label{GLR}
    \frac{{\rm d} \Gamma_1(Z^*\to\ell^+\ell^-)}
    {{\rm d} \cos\theta} \propto
    \vert d^1_{00}\vert\,^2\sim\cos^2\theta,
\end{equation}
than the previously considered ones. A striking feature of the
distribution is the forbidden decay direction perpendicular to the
beam. It leads to a profound dip at $\cos\theta=0$~\cite{proposal}.
The same dip presents also at $\Delta\eta=0$~\cite{deta} (the solid
curve in Fig.~\ref{fig:angular}):
\begin{equation}\label{dy1star}
    \frac{1}{\Gamma^*_1}\frac{{\rm d}\Gamma^*_1}{{\rm d}\Delta\eta}=
    \frac{6{\rm e}^{\Delta\eta}({\rm e}^{\Delta\eta}-1)^2}{({\rm e}^{\Delta\eta}+1)^4}.
\end{equation}

It can be seen from Fig.~\ref{fig:angular} that the excited bosons
have unique signature in the angular distribution. They manifest
themselves through the absolute minimum at $\Delta\eta=0$ and
absolute maximum at $\Delta\eta=\ln(3+\sqrt{8})\approx 1.76$\,. The
latter corresponds to the polar angle $\theta=45^\circ$ and a little
bit contradicts to the common opinion about an expected signal from
new physics.

The spin-2 resonances, like Kaluza--Klein excitations, lead to the
following decay distributions depending on initial parton
configurations
\begin{equation}\label{gg}
    \frac{{\rm d} \Gamma_{gg}(G^*\to\ell^+\ell^-)}
    {{\rm d} \cos\theta} \propto \vert d^{2}_{2,1}\vert\,^2
    +\vert d^{2}_{2,-1}\vert\,^2\sim 1-\cos^4\theta
\end{equation}
and
\begin{eqnarray}\label{qq}
    \frac{{\rm d} \Gamma_{q\bar{q}}(G^*\to\ell^+\ell^-)}
    {{\rm d} \cos\theta} &\propto& \vert d^{2}_{1,1}\vert\,^2
    +\vert d^{2}_{1,-1}\vert\,^2\nonumber\\
    &\sim&
    1-3\cos^2\theta+4\cos^4\theta.
\end{eqnarray}
Taking into account their corresponding weights, 3/5 and 2/5, for
the inclusive process $pp\to G^*\to\ell^+\ell^-$ \cite{KK} and using
eq. (\ref{trans}) one can obtain the distribution on $\Delta\eta$
(the dashed curve in Fig.~\ref{fig:angular})
\begin{equation}\label{dy2}
    \frac{1}{\Gamma_2}\frac{{\rm d}\Gamma_2}{{\rm d}\Delta\eta}=
    \frac{2{\rm e}^{\Delta\eta}({\rm e}^{4\Delta\eta}+18{\rm e}^{2\Delta\eta}+1)}
    {({\rm e}^{\Delta\eta}+1)^6},
\end{equation}
which also peaks at $\Delta\eta=0$ as in the most exotic models.

\section{Cut optimization for excited bosons}

From Fig.~\ref{fig:angular} one can see that all distributions
besides of the excited bosons peak at $\Delta\eta=0$. Since the main
background comes from the $\gamma/Z$ bosons, which have the minimal
gauge couplings with quarks and lepton, they also populate the
central pseudorapidity region. Therefore, it is difficult to
disentangle new physics signal using this distribution unless in the
case of the excited bosons. Applying the proper cut $\Delta\eta > a$
one can suppress unwanted background contribution and enhance the
signal significance, ${\cal S} = s/\sqrt{b}$, choosing its maximum.

\begin{figure}[t]
\epsfig{file=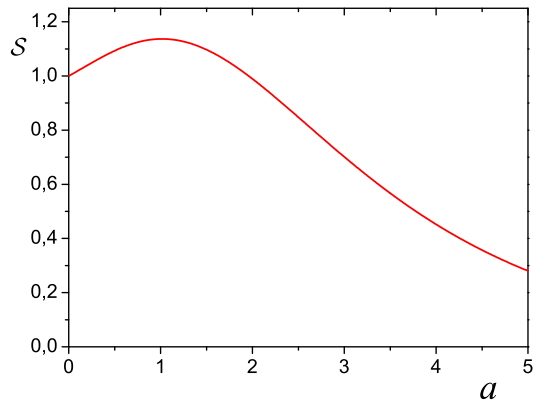,width=0.5\textwidth}
\caption{\label{fig:sig} The relative signal significance
distribution of the excited bosons as functions of the cut parameter
$a$ on $\Delta\eta$.}
\end{figure}

So, we can define the relative significance of $Z^*$ signal to
$Z'$-like background as the ratio of the definite integral from $a$
to infinity of the normalized signal distributions (\ref{dy1star})
to the square root of the SM background, dominated by distribution
(\ref{dy1prime}),
\begin{equation}\label{relSig}
    {\cal S}(a) =
    \frac{\int_a^\infty \frac{1}{\Gamma_1^*}
    \frac{{\rm d}\Gamma_1^*}{{\rm d}\Delta\eta}{\rm d}\Delta\eta}
    {\sqrt{\int_a^\infty\frac{1}{\Gamma'_1}
    \frac{{\rm d}\Gamma'_1}{{\rm d}\Delta\eta}{\rm d}\Delta\eta}}=
    \frac{1-\left(\frac{{\rm e}^a-1}{{\rm e}^a+1}\right)^3}
    {\sqrt{1-\frac{{\rm e}^{3a}-1}{({\rm e}^a+1)^3}}}
\end{equation}
The distribution reaches the maximum value around 1.14 at $a \approx
1.02$ (Fig.~\ref{fig:sig}).

The corresponding cut at the maximum signal significance suppresses
around 38\% of the background and only 10\% of the signal. This
leads to relative enhancement in sensitivity, which is equivalent to
adding approximately 29\% of data in comparison with usual
sensitivity without any cuts.

\section{Conclusions}

In this paper we have presented $\Delta\eta$-distributions for all
possible resonanses with spin up to 2. On this basis we have
proposed the novel optimization of angular distribution cut aimed at
the most effective search for the new resonances with different
angular distributions in dilepton events. In particular, this allows
to enhance the significance of bump search for the excited bosons.

\section*{Acknowledgements}
The work of M.V. Chizhov was partially supported by the grant of
Plenipotentiary of the Republic of Bulgaria in JINR for 2011 year.


\end{document}